\documentclass[conference]{IEEEtran}
\makeatletter
\def\ps@headings{%
\def\@oddhead{\mbox{}\scriptsize\rightmark \hfil \thepage}%
\def\@evenhead{\scriptsize\thepage \hfil \leftmark\mbox{}}%
\def\@oddfoot{}%
\def\@evenfoot{}}
\makeatother
\usepackage{graphicx}
\usepackage{dcolumn}
\usepackage{bm}
\usepackage{epsfig}
\usepackage{psfrag}
\usepackage{subfigure}
\usepackage{color}
\usepackage{amsmath, amssymb, latexsym,amsfonts,epsfig, graphicx, verbatim}
\usepackage[nospace]{cite}
\usepackage{algorithmic}
\usepackage{algorithm}

\newcommand{\eq}{\!\!=\!}
\newcommand{\m}{\!-\!}
\newcommand{\Prob}{\mathbb{P}}
\newcommand{\Mean}{\mathbb{E}}

\newcommand{\est}[1]{\widehat{#1}}

\begin{document}

\title{On the bias of BFS}

\author{\IEEEauthorblockN{Maciej Kurant}
\IEEEauthorblockA{School of Computer \& Comm. Sciences\\
EPFL, Lausanne, Switzerland\\
{\em maciej.kurant@gmail.com}}
\and
\IEEEauthorblockN{Athina Markopoulou}
\IEEEauthorblockA{EECS Dept\\
University of California, Irvine\\
{\em athina@uci.edu}}
\and
\IEEEauthorblockN{Patrick Thiran}
\IEEEauthorblockA{School of Computer \& Comm. Sciences\\
EPFL, Lausanne, Switzerland\\
{\em patrick.thiran@epfl.ch}}
}

\maketitle

\begin{abstract}
Breadth First Search (BFS) is widely used for measuring large unknown graphs, such as Online Social Networks.
It has been empirically observed that an incomplete BFS is biased toward high degree nodes.
In contrast to more studied sampling techniques, such as random walks, the precise bias of BFS has not been characterized to date.

In this paper, we quantify the degree bias of BFS sampling. In particular, we calculate the node degree distribution expected to be observed by BFS as a function of the fraction of covered nodes, in a random graph $RG(p_k)$ with a given degree distribution~$p_k$.
Furthermore, we also show that, for $RG(p_k)$, all commonly used graph traversal techniques (BFS, DFS, Forest Fire, and Snowball Sampling) lead to the same bias, and we show how to correct for this bias.
To give a broader perspective, we compare this class of exploration techniques to random walks that are well-studied and easier to analyze.
Next, we study by simulation the effect of graph properties not captured directly by our model.
We find that the bias gets amplified in graphs with strong positive assortativity.
Finally, we demonstrate the above results by sampling the Facebook social network, and we provide some practical guidelines for graph sampling in practice.
\end{abstract}

\section{Introduction}

A large body of work in the networking community focuses on topology measurements at various levels, including the Internet, the Web (WWW), peer-to-peer (P2P) and online social networks (OSN).
The size of these networks and other practical restrictions make measuring the entire graph impossible. Instead, researchers typically collect and study a small but ``representative'' sample.
In this paper, we are particularly interested in sampling networks that naturally allow to explore the neighbors of a given node (which is the case in WWW, P2P and OSN). A number of graph exploration techniques
use this basic operation for sampling. They can be roughly classified in two categories: (a)~with replacement (random walks), and (b)~without replacement (graph traversal techniques).

In the first category, random walks, nodes can be revisited. This category includes the classic Random Walk (RW) as well as the Metropolis-Hastings Random Walk (MHRW). They are used for sampling of nodes on the Web~\cite{monica}, P2P networks~\cite{willinger-unbiased-p2p,Rasti09-RDS_Characterizing_Unstructured_Overlays, gkantsidis04randomwalksp2p}, OSNs~\cite{Gjoka09_Facebook_arxiv, Twitter08} and large graphs in general~\cite{Leskovec06_sampling-largegraphs}.
Random walks are well studied~\cite{Lovasz93} and result in samples that have either no bias (MHRW) or a known bias (RW) that can be corrected for.
Random walks are {\em not the focus} of this paper, but are discussed as baseline for comparison.

\begin{figure}[t]
\psfrag{y0}[r][c][0.9]{$\langle k^*\rangle$ - }
\psfrag{y1}[l][c][0.8]{expected observed}
\psfrag{y2}[l][c][0.8]{average node degree}
\psfrag{k3}[r][c][0.9]{$\langle k\rangle$}
\psfrag{k2}[r][c][0.9]{$\frac{\langle k^2\rangle}{\langle k\rangle}$}
\psfrag{i}[c][c][0.9]{$f$ - fraction of sampled nodes}
\psfrag{tot}[c][c][0.9]{1}
\psfrag{rw}[c][c][0.7]{Random Walk (RW)}
\psfrag{a}[l][c][0.7]{Graph traversal techniques:}
\psfrag{b}[l][c][0.7]{- BFS}
\psfrag{c}[l][c][0.7]{- DFS}
\psfrag{d}[l][c][0.7]{- Forest Fire}
\psfrag{e}[l][c][0.7]{- Snowball}
\psfrag{mhrw}[c][c][0.7]{Metropolis-Hastings Random Walk (MHRW)}
\includegraphics[width=0.49\textwidth]{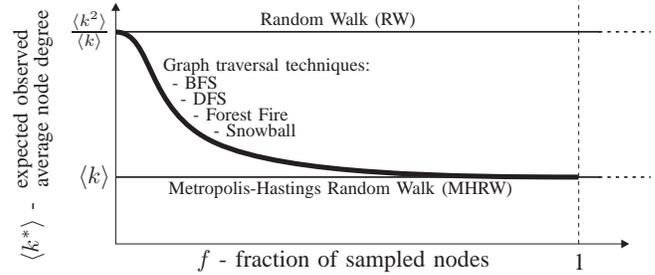}
\caption{{\bf Overview of results.} We calculate the average node degree $\langle k^*\rangle$ (and the full degree distribution, not shown) expected to be observed by BFS in a random graph $RG(p_k)$ with a given degree distribution $p_k$, as a function of the fraction of sampled nodes $f$. We show RW and MHRW as a reference.
$\langle k\rangle$ is the real average node degree, and $\langle k^2\rangle$ is the real average squared node degree.
\quad~{\em Observations:}
\quad~(1)~For a small sample size, BFS has the same bias as RW; with increasing $f$, the bias decreases; a complete BFS ($f\eq 1$) is unbiased,  as is MHRW (or uniform sampling).
\quad~(2)
All common graph traversal techniques (that do not revisit the same node) lead to the same bias.
\quad~(3)~The shape of the BFS curve depends on the real node degree distribution~$p_k$, but it is always monotonically decreasing.
}
\vspace{-0.4cm}
\label{fig:contributions}
\end{figure}

In the second category, graph traversal techniques, each node is visited exactly once (if we let the process run until completion). These methods vary in the order in which they visit the nodes; examples include BFS, Depth-First Search (DFS), Forest Fire (FF) and Snowball Sampling (SBS).
Graph traversals, especially BFS, are very popular and widely used for sampling large networks, {\em e.g.} WWW~\cite{Najork01} 
or OSNs ~\cite{Ahn-WWW-07, Mislove-IMC-07, Wilson09}.
One reason is that BFS is well-known (a textbook technique) and easy to understand. Another reason is that (incomplete) BFS
collects a full view (all nodes and edges) of some particular region in the graph, which is sometimes believed to be representative of the entire graph. {\em E.g.,} a BFS sample of a lattice is a (smaller) lattice.

Unfortunately, this intuition often fails.
It was observed empirically that BFS introduces a bias towards high-degree nodes~\cite{Najork01,Lee-Phys-Rev-06,snowball-bias}. We also confirmed this fact in a recent measurement of Facebook~\cite{Gjoka09_Facebook_arxiv}, where our BFS crawler found the average node degree $\langle k^{\scriptscriptstyle BFS}\rangle\simeq 324$, while the real value is only $\langle k\rangle\simeq 94$, {\em i.e.,} about 3.5 times smaller!



Given the popularity of BFS on one hand, and its bias on the other hand, it is surprising that we still know relatively little about the statistical properties of node sequences returned by BFS. Indeed, sampling without replacement introduces complex dependencies, no rigorous analytical explanation of the observed biases of BFS was available to date.

Our work is a first step toward understanding the statistical characteristics of incomplete BFS sampling.
In particular, we calculate precisely the node degree distribution expected to be observed by BFS as a function of the fraction of sampled nodes in a random graph $RG(p_k)$ with a given (and arbitrary) degree distribution $p_k$. We accompany this central result with additional related contributions. First, we show that in $RG(p_k)$, BFS is equivalent to other graph traversal techniques, such as Depth First Search (DFS), Snowball Sampling, and Forest Fire (FF).
%
Second, we compare the bias of BFS (and other traversal techniques) to that of random walks.
As shown in Fig.~\ref{fig:contributions} and as also formally demonstrated in this paper, in the beginning of the exploration process, BFS exhibits exactly the same bias as the Random Walk~(RW). With increasing fraction of sampled nodes~$f$, this bias monotonically decreases. When the BFS is complete ($f=1$), there is no bias, as it is can also be achieved by the Metropolis-Hastings Random Walk (MHRW). Moreover, given a biased sample, we derive an unbiased estimator of the original node degree distribution.

In addition, we use simulation to confirm our analysis and investigate the effect of graph properties, such as assortativity, not captured directly by~$RG(p_k)$. We complement it with real-world measurements of the Facebook social network.

{\em Scope.}
Our theoretical results hold for the random graph model $RG(p_k)$ described in Section~\ref{sec:Graph model}. We study some extensions of this model in simulations in Section~\ref{sec:Simulation}. We also restrict our attention to BFS sampling of static graphs.

The outline of the paper is as follows. Section \ref{sec:Related_Work} discusses related work. Section \ref{sec:Algorithms} presents the graph sampling algorithms under study. Section \ref{sec:Graph model} presents the random graph model used in this paper. Section \ref{sec:Analysis} analyzes the expected degree distribution of various graph sampling techniques; in particular the main results related to BFS are derived in Section~\ref{sec:Analysis}.B.
Section~\ref{sec:Correcting for node degree bias} shows how to correct for the bias.
Section~\ref{sec:Simulation} presents simulation results. Section~\ref{sec:Facebook}, demonstrates the above ideas by sampling a real world network, Facebook, and provides  hints for graph sampling in practice. Section~\ref{sec:Conclusion} concludes and outlines future work.

\section{Related Work}\label{sec:Related_Work}


{\em BFS used in practice.} BFS is widely used today for exploring large networks, such as OSNs. The following list provides some examples but is by no means exhaustive. In \cite{Ahn-WWW-07}, Ahn et al.   used BFS to sample Orkut and MySpace. In \cite{Mislove-IMC-07} and \cite{MisloveWosn08}, Mislove et al.  used BFS  to crawl the social graph in four popular OSNs: Flickr, LiveJournal, Orkut, and YouTube. In \cite{Wilson09}, Wilson et al. measured the social graph and the user interaction graph of Facebook using several BFSs, each BFS constrained in one of the largest 22 regional Facebook networks. In our recent work \cite{Gjoka09_Facebook_arxiv}, we have also crawled Facebook using various sampling techniques, including BFS, RW and MHRW.
 It has been empirically observed that incomplete BFS and its variants introduce bias towards high-degree nodes \cite{Najork01,Lee-Phys-Rev-06,snowball-bias}. We also confirmed this in Facebook \cite{Gjoka09_Facebook_arxiv}, an observation  that in fact inspired this paper.

{\em Analyzing BFS.} 
To the best of our knowledge, the sampling bias of BFS has not been analyzed so far.
\cite{Kim06_poisson_cloning} and \cite{Achlioptas05_On_the_bias_of_traceroute_sampling} are the closest related papers to our methodology. The original paper by Kim~\cite{Kim06_poisson_cloning} analyzes the size of the largest connected component in classic Erd\"os-R\'enyi random graph by essentially applying the configuration model with node degrees chosen from a Poisson distribution. 
To match the stubs (or `clones' in~\cite{Kim06_poisson_cloning}) uniformly at random in a tractable way, Kim proposes a ``cut-off line'' algorithm: he first assigns each stub a random index from $[0,np]$, and next progressively scans this interval. 
Achlioptas et al. used this powerful idea in~\cite{Achlioptas05_On_the_bias_of_traceroute_sampling} to study the bias of traceroute sampling in random graphs with a given degree distribution.
The basic operation in \cite{Achlioptas05_On_the_bias_of_traceroute_sampling} is traceroute ({\em i.e.,} ``discover a path'') and is performed from a single node to all other nodes in the graph. 
The union of the observed paths forms a ``BFS-tree'', which includes all nodes but misses some edges (\emph{e.g.,} those between nodes at the same depth in the tree). In contrast, the basic operation in the traversal methods presented in our paper is to discover all neighbors of a node, and it is applied to all nodes in increasing distance from the origin. 
Another important difference 
is that~\cite{Achlioptas05_On_the_bias_of_traceroute_sampling} studies a completed BFS-tree, whereas we study the sampling process when it has visited only a fraction $f<1$ of nodes; a completed BFS ($f\eq 1$) is trivial in our case (it has no bias).

There is also a large body of literature on unequal probability sampling without replacement ~\cite{Shahbaz03_Sampling_with_Unequal_Probabilities}.
Although, at first, it seems to be a promising path to follow, to the best of our knowledge, none of the existing results is directly applicable to our problem.
This is because, speaking in the terms used later in this paper, the available results either (i)~require the knowledge of $q_k(f)$ as an input, or (ii)~propose how to calculate $q_k(f)$ for the first two nodes only.

Another recent paper related to BFS bias is ~\cite{Illenberger09_snowball_bias_correction}. The paper is about Snowball Sampling~\cite{Goodman61_Snowball_sampling}, which is similar to BFS, and proposes a heuristic approach to correct the degree biases in $i$th generation of Snowball based on the values measured in generation~$i\m 1$. The authors show by simulation that this technique performs moderately well, especially when a significant fraction of nodes have been covered.

{\em Random Walks.} Simple and metropolized random walks  are also used for crawling OSNs \cite{Gjoka09_Facebook_arxiv,Twitter08}, P2P networks \cite{willinger-unbiased-p2p,Rasti09-RDS_Characterizing_Unstructured_Overlays, gkantsidis04randomwalksp2p}, the web \cite{monica} and large graphs in general \cite{Leskovec06_sampling-largegraphs}. Random walks are well-studied~\cite{Lovasz93},
their bias is known and can be corrected. Random walks are {\em not} the focus of the paper but are used as baseline for comparison. 

\section{Graph exploration techniques}\label{sec:Algorithms}
Let $G=(V,E)$ be a connected graph with the set of vertices $V$, and a set of undirected edges $E$.
Initially, $G$ is unknown, except for one (or some limited number of) seed node(s).
When sampling through graph \emph{exploration}, we begin at the seed node, and we recursively visit (one, some or all) of its neighbors.
We distinguish two main categories of exploration techniques: with and without replacement.

\subsection{Exploration with replacement (random walks)}\label{subsec:Exploration with replacements}
Exploration \emph{with replacement}, or simply a \emph{walk}, allows revisiting the same node many times. Consider the following classic examples:

\smallskip
\subsubsection{Random Walk (RW)}
In this classic sampling technique~\cite{Lovasz93}, we start at some seed node. At every iteration, the next-hop node~$v$ is chosen uniformly at random among the neighbors of the current node~$u$. It is easy to see that RW introduces a linear bias towards nodes of high degree~\cite{Lovasz93}.

\smallskip
\subsubsection{Metropolis Hastings Random Walk (MHRW)}

In this technique, as in RW, the next-hop node~$w$ is chosen uniformly at random among the neighbors of the current node~$u$. However, with a probability that depends on the degrees of $w$ and~$u$, MHRW performs a self-loop instead of moving to $w$.
More specifically, the probability $P_{u,w}$ of moving from $u$ to $w$ is as follows~\cite{mcmc-book}:
\begin{equation}\label{eq:P_u,w}
    P_{u,w} = \left\{ \begin{array}{ll}
\frac{1}{k_u} \cdot \min(1, \frac{k_u}{k_w}) & \textrm{if $w$ is a neighbor of $u$,} \\
1- \sum_{y\neq u} P_{u,y} & \textrm{if $w=u$,} \\
0 & \textrm{otherwise},
\end{array} \right.
\end{equation}
where $k_v$ is the degree of node $v$. Essentially, MHRW reduces the transitions to high degree nodes and thus eliminates the degree bias of RW. This property of MHRW was recently exploited in various network sampling contexts \cite{willinger-unbiased-p2p,Twitter08,Gjoka09_Facebook_arxiv,Rasti09-RDS_Characterizing_Unstructured_Overlays}.

\smallskip
\subsubsection{Respondent-Driven Sampling (RDS)}
RDS was proposed and studied in the field of social sciences to penetrate hidden populations, such as that of drug addicts~\cite{Heckathorn97_RDS_introduction,Salganik04_RDS}. In the network sampling terminology, at each iteration RDS selects randomly exactly $n$~neighbors (typically $n\simeq 3$) of the current node~$u$ and schedules them to visit later.
RDS visits the nodes in the order they were scheduled.
Thus, RDS is a modification of Snowball Sampling (described below) that \emph{allows node revisiting}.
\footnote{In practical RDS surveys in human populations, nodes (people) are not revisited. However, the revisiting assumption is necessary to formally correct for the degree bias~\cite{Salganik04_RDS}. The authors of \cite{Salganik04_RDS} argue that this approximation is valid if the sample size is relatively small compared to the population size. In this paper we formally confirm this claim.}
RDS introduces a degree bias that is known and can be corrected for. It was demonstrated in~\cite{Salganik04_RDS} on the example with $n\eq 1$, which reduces RDS precisely to Random Walk (RW). This approach was recently tested in~\cite{Rasti09-RDS_Characterizing_Unstructured_Overlays} on various graph models and unstructured P2P networks.


\subsection{Exploration without replacement (graph traversals)}
In contrast, exploration \emph{without replacement}, or \emph{graph traversal}, never revisits the same node and. At the end of the process, and assuming that the graph is connected, all nodes are visited.

\smallskip
\subsubsection{Breadth First Search (BFS)}
BFS is a classic graph traversal algorithm that starts from the seed and progressively explores all neighbors. At each new iteration the {\em earliest} explored but not-yet-visited node is selected next. Thus, BFS discovers all nodes within some distance from the seed.

\smallskip
\subsubsection{Depth First Search (DFS)}
This technique is similar to BFS, except that at each iteration we select the latest explored but not-yet-visited node. As a result, DFS explores first the nodes that are faraway (in the number of hops) from the seed.

\smallskip
\subsubsection{Forest Fire (FF)}\label{sec:Forest Fire (FF)}
FF is a randomized version of BFS, where for every neighbor $v$ of the current node, we flip a coin, with probability of success $p$, to decide if we explore $v$. FF reduces to BFS for $p\eq 1$. It is possible that this process dies out before it covers all nodes. In this case, in  order to make FF comparable with other techniques, we revive the process from a random node already in the sample. Forest Fire is inspired by the graph growing model of the same name proposed in~\cite{Leskovec05_Forest_Fire} and is used as a graph sampling technique in~\cite{Leskovec06_sampling-largegraphs}.

\smallskip
\subsubsection{Snowball Sampling (SBS)}
Snowball Sampling is a precursor of RDS and a term loosely used for BFS-like traversal techniques.
According to a classic definition  by Goodman~\cite{Goodman61_Snowball_sampling}, an $n$-name Snowball Sampling is  similar to BFS, but at every node $v$, not all $k_v$, but \emph{exactly} $n$ neighbors are chosen randomly out of all $k_v$ neighbors of $v$. These $n$ neighbors are scheduled to visit, but only if they have not been visited before.

\section{Graph model $RG(p_k)$}\label{sec:Graph model}

\begin{table}
  \centering
{\footnotesize
\begin{tabular}{l|l}
  \hline
  $G=(V,E)$ & graph $G$ with nodes $V$ and edges $E$\\
  $k_v$ & degree of node $v$\\
  $p_k \ = \frac{1}{|V|}\sum_{v\in V} 1_{k_v=k}$ & degree distribution in $G$\\
  $q_k$ & expected observed degree distribution\\
  $\est{q}_k$ & observed degree distribution\\
  $\est{p}_k$ & estimated original degree distribution in $G$\\
  $\langle k\rangle \ = \sum_k k\, p_k$ & average node degree in $G$ \\
  $\langle k^*\rangle \ = \sum_k k\, q_k $ & expected observed average node degree\\
  $f$   &  fraction of nodes covered by the sample\\
  \hline
\end{tabular}
}
\caption{Notation Summary. `Observed' means calculated directly from the sample.}
\label{Tab:notation}\vspace{-0.8cm}
\end{table}

A basic important graph property is the node degree distribution $p_k$, {\em i.e.}, the fraction of nodes with degree equal to~$k$, for all $k\geq0$.\footnote{As we define $p_k$ as a `fraction', not the `probability', $p_k$ determines the degree sequence in the graph, and vice versa.}
Depending on the network, the degree distribution can vary, ranging from constant-degree (in regular graphs), a distribution concentrated around the average value ({\em e.g.,} in Erd\"os-R\'enyi random graphs or in well-balanced P2P networks), to heavily right-skewed distributions with $k$ covering several decades (in WWW, unstructured P2P, Internet at the Autonomous System level, OSNs).
We handle all these cases by assuming that we are given \emph{any} fixed node degree distribution $p_k$.
Other than that, the graph $G$ is completely random. That is, $G$ is drawn uniformly at random from the set of all multigraphs\footnote{A multigraph is a graph that accepts multiple edges and self-loops.
}
with degree distribution $p_k$. We denote this model by $RG(p_k)$.

We use a classic technique to generate $RG(p_k)$, called \emph{configuration model}~\cite{Molloy95,Newman03_Review}: each node~$v$ is given $k_v$ ``stubs'' (or ``edges-to-be'').
Next, all these $\sum_{v\in V} k_v= 2|E|$ stubs are randomly matched in pairs, until all stubs are exhausted (and $|E|$ edges are created).
In Fig.~\ref{fig:stubs_on_interval} (ignore the rectangular interval [0,1] for now), we present four nodes with their stubs (left) and an example of their random matching (right).

\section{\label{sec:Analysis}Analyzing the Node Degree Bias}
In this section, we study the node degree bias observed when the graph exploration techniques of Section~\ref{sec:Algorithms} are run on the random graph $RG(p_k)$ of Section~\ref{sec:Graph model}. In particular, we derive the node degree distribution $q_k$ and the average node degree~$\langle k^*\rangle$ expected to be observed, as a function of the original degree distribution $p_k$ and, in the case of BFS, of the fraction of sampled nodes $f$.

\subsection{Exploration with replacement (walks)}
We begin by summarizing the relevant results known for walks, in particular for RW and MHRW. They will serve as a reference point for our main analysis of graph traversals in the next section.

\subsubsection{Random Walk (RW)} \label{Weighted sampling with replacements}

Random walk have been widely studied; see \cite{Lovasz93} for an excellent survey. In any given connected and aperiodic graph, the probability of being at a particular node $v$ converges at equilibrium to the stationary distribution~$\pi_v \eq \frac{k_v}{2|E|}$.
Therefore, the expected observed degree distribution $q_k$ is
\begin{eqnarray}
\nonumber   q_k &=&  \sum_v \pi_v \cdot 1_{\{k_v=k\}} \ =\ \frac{k}{2|E|}\cdot \sum_v 1_{\{k_v=k\}}\ = \\
\label{eq:q_k_RW}   &=& \frac{k}{2|E|}\, p_k\,|V| \ =\ \frac{k\, p_k}{\langle k\rangle},
\end{eqnarray}
where $\langle k\rangle$ is the average node degree in~$G$. Eq.~(\ref{eq:q_k_RW}) is essentially similar to calculation for RDS in \cite{Newman01_EgoCentered_Networks,Salganik04_RDS}.
As this holds for any fixed (and connected and aperiodic) graph, it is also true for all connected graphs generated by the configuration model.
Consequently, the expected observed average node degree is
\begin{equation}\label{eq:Ek_RW}
    \langle k^*\rangle = \frac{\sum_k k^2\, p_k}{\langle k\rangle}\ =\ \frac{\langle k^2\rangle}{\langle k\rangle},
\end{equation}
where $\langle k^2\rangle$ is the average squared node degree in $G$. We show this value $\frac{\langle k^2\rangle}{\langle k\rangle}$ in Fig.~\ref{fig:contributions}.


%
%

\subsubsection{Metropolis Hastings Random Walk (MHRW)}
It is easy to show that the transition matrix~$P_{u,w}$ shown in Eq.(\ref{eq:P_u,w}) leads to a uniform stationary distribution~$\pi_v \eq \frac{1}{|V|}$~\cite{mcmc-book}, and consequently:
\begin{eqnarray}
\label{eq:q_k_UNI}  q_k &=& p_k \\
\label{eq:Ek_UNI}  \langle k^*\rangle &=& \sum_k k\cdot p_k\ =\ \langle k\rangle.
\end{eqnarray}
In Fig.~\ref{fig:contributions}, we show that MHRW estimates the true mean.

\subsection{Exploration without replacement (Main Result)}

In both RW and MHRW the nodes can be revisited. So the state of the system at iteration~$i\!+\!1$ depends only on iteration~$i$, which makes it possible to analyze as Markov Chains. In contrast, graph traversals do not allow for node revisits, which introduces crucial dependencies between all the iterations and significantly complicates the analysis.
To handle these dependencies, we adopt an elegant technique recently introduced in~\cite{Kim06_poisson_cloning} (to study the size of the largest connected component) and extended in~\cite{Achlioptas05_On_the_bias_of_traceroute_sampling} (to study the bias of traceroute sampling). However, our work differs in many aspects from both \cite{Kim06_poisson_cloning} and \cite{Achlioptas05_On_the_bias_of_traceroute_sampling}, which we comment in detail in the related work Section~\ref{sec:Related_Work}.

\smallskip
\subsubsection{Exploration without replacement at the stub level}

We begin by defining Algorithm~1 (below) - a general graph traversal technique that collects a sequence of nodes $S$, without replacements. To be compatible with the configuration model (see Section~\ref{sec:Graph model}), we are interested in the process \emph{at the stub level}, where we consider one stub at a time, rather than one node at a time. An integral part of the algorithm is a queue $Q$, that keeps the discovered, but still not-yet-followed stubs. We start the algorithm by adding to $Q$ all the stubs of some initial node~$v_1$, and by setting $S\eq[v_1]$. Next, at every iteration, we pop one stub $a$ from $Q$, and follow it to discover its partner-stub~$b$, and $b$'s owner $v(b)$. If node $v(b)$ is not yet discovered, {\em i.e.,} if $v(b)\notin S$, then we append $v(b)$ to $S$ and we add to $Q$ all other stubs of $v(b)$. More formally:
%
\begin{algorithm}[h!]
\caption{Stub-Level Graph Traversal}
\label{alg:1}
\begin{algorithmic}[1]
\STATE $S\gets [v_1]$ \ and \ $Q\gets$ [all stubs of $v_1$] 
\WHILE {$Q$ is nonempty}
    \STATE Pop $a$ from $Q$
    \STATE Discover $a$'s partner $b$
    \IF {$v(b)\notin S$}                       
            \STATE Append $v(b)$ to $S$     
            \STATE Add to $Q$ all stubs of $v(b)$ except $b$
    \ELSE {}
    \STATE Remove $b$ from $Q$
    \ENDIF
\ENDWHILE
\end{algorithmic}
\end{algorithm}

Depending on the scheduling discipline for the elements in~$Q$ (line~3), Algorithm~1 implements BFS (for a first-in first out scheduling), DFS (last-in first-out) or Forest Fire (first-in first-out with randomized stub losses). Line~9 guarantees that the algorithm never tracebacks the edges, {\em i.e.,} that stub $a$ popped from $Q$ in line~3 never belongs to an edge that has already been traversed in the opposite direction.

\begin{figure*}[t]
\psfrag{time}[c][c][1]{time $t$ (index)}
\psfrag{t}[c][c][0.8]{current time $t$}
\psfrag{v1}[c][c][0.9]{$v_1$}
\psfrag{v2}[c][c][0.9]{$v_2$}
\psfrag{v3}[c][c][0.9]{$v_3$}
\psfrag{v4}[c][c][0.9]{$v_4$}
\includegraphics[width=1\textwidth]{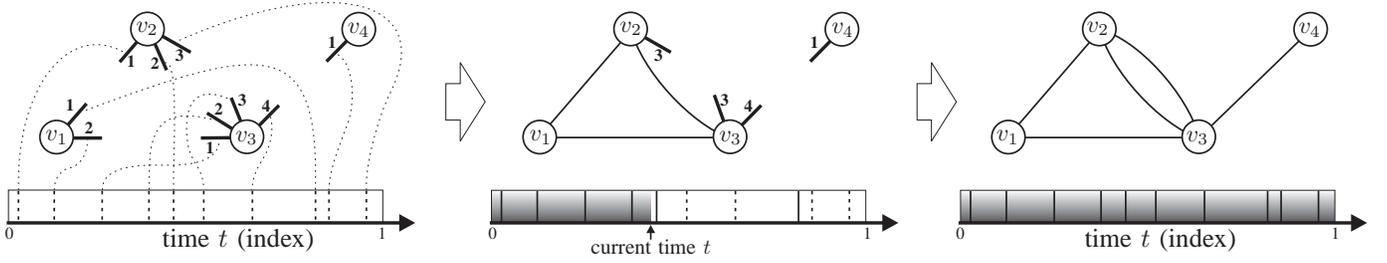}
\caption{An illustration of the stub-level, on-the-fly graph exploration without replacements. In this particular example, we show an execution of BFS starting at node $v_1$.
\quad \textbf{Left:} Initially, each node $v$ has $k_v$ stubs, where $k_v$ is a given target degree of $v$. Each of these stubs is assigned a real-valued number drawn uniformly at random from the interval $[0,1]$ shown below the graph. Next, we follow Algorithm~1 with a starting node $v_1$. The numbers next to the stubs of every node $v$ indicate the order in which these stubs are added to the queue~$Q$.
\quad \textbf{Center:} The state of the system at time $t$. All stubs in $[0,t]$ have already been matched (the indices of matched stubs are set in plain line). All unmatched stubs are distributed uniformly at random on $(t,1]$. This interval can contain also some (here two) already matched stubs.
\quad \textbf{Right:} The final result is a realization of a random graph $G$ with a given node degree sequence (i.e., of the configuration model). $G$ may contain self-loops and multiedges.
}
\label{fig:stubs_on_interval}
\vspace{-0.4cm}
\end{figure*}

\smallskip
\subsubsection{Discovery on the fly}
In line~4 of Algorithm 1, we follow stub $a$ to discover its partner $b$.
In a fixed graph~$G$, this step is deterministic. In the configuration model~$RG(p_k)$, 
a fixed graph~$G$ is obtained by matching all the stubs uniformly random. Next we can sample this fixed graph and average it over the space of all the random graphs~$RG(p_k)$ that have just been constructed. Unfortunately, this quickly leads to complex combinatorial problems. We adopt therefore an alternative
and more tractable construction of a fixed graph with is an iterative sampling from the set of random graphs, by selecting~$b$ `on the fly' (i.e, every time line~4 is executed), uniformly at random from all the unmatched stubs. By the principle of deferred decisions~\cite{Randomized_Algorithms_book}, these two approaches are equivalent.




\smallskip
\subsubsection{Breaking the dependencies}\label{subsec:Breaking the dependencies}
There is still one problem with the `on the fly' method. It selects stub~$b$ uniformly at random from all the \emph{unmatched} stubs. This introduces dependencies between the stubs and across all the iterations.
We remedy this by implementing the `on the fly' approach as follows.
First, we assign each stub a real-valued index $t$ drawn uniformly at random from the interval $[0,1]$.
Then, every time we process line~4, we pick $b$ as the unmatched stub with the smallest index. We can interpret this as a continuous-time process, where we determine progressively the partners of stubs popped from queue~$Q$, by scanning the interval from `time' $t\eq 0$ to $t\eq 1$ in a search of unmatched stubs. Because the indices chosen by the stubs are independent from each other, the above trick breaks the dependence between the stubs, which is a crucial for making this approach tractable.
%
%

In Fig.~\ref{fig:stubs_on_interval}, we present an example execution of Algorithm~1, where line~4 is implemented as described above.

\smallskip
\subsubsection{Expected sampled degree distribution $q_k$}
Now we are ready to derive the expected observed degree distribution $q_k$. Recall that all the stub indices are chosen independently and uniformly from $[0,1]$.
A vertex~$v$ with degree~$k$ is not sampled yet at time $t$ if the indices of all its $k$ stubs are larger than~$t$, which happens with probability~$(1\m t)^k$.
So the probability that $v$ is sampled before time $t$ is $1\m (1\m t)^k$. Therefore, the expected fraction of vertices of degree $k$ sampled before $t$ is
\begin{equation}\label{eq:f_k(t)}
    f_k(t) = p_k  (1\m (1\m t)^k).
\end{equation}
By normalizing (\ref{eq:f_k(t)}), we obtain the expected observed (sampled) degree distribution at time $t$:
\begin{equation}\label{eq:q_k_t}
   q_k(t)\ =\  \frac{f_k(t)}{\sum_l f_l(t)}\ =\ \frac{p_k (1 - (1\m t)^k)}{\sum_l p_l (1 - (1\m t)^l)}.
\end{equation}
Unfortunately, it is difficult to interpret $q_k(t)$ directly, because~$t$ is proportional neither to the number of matched edges nor to the number of discovered nodes. Recall that our primary goal is to express $q_k$ as a function of fraction $f$ of covered nodes.
We achieve this by calculating $f(t)$ - the expected fraction of nodes, of any degree, visited before time $t$
\begin{equation}\label{eq:f(t)}
    f(t) =  \sum_k f_k(t) =  1 - \sum_k p_k (1\m t)^k
\end{equation}
Because $p_k\geq 0$, and $p_k> 0$ for at least one $k>0$, the term $\sum_k p_k (1\m t)^k$ is  continuous and strictly decreasing from 1 to 0 with $t$ growing from 0 to 1. Thus, for $f\in [0,1]$ there exists a well defined function $t(f)$  that satisfies Eq.(\ref{eq:f(t)}), {\em i.e.,} the inverse of $f(t)$. Although we cannot compute $t(f)$ analytically (except in some special cases such as for $k\leq 4$), it is straightforward to find it numerically. Now, we can rewrite Eq.(\ref{eq:q_k_t}) as
\begin{equation}\label{eq:q_k_f}
   q_k(f)\ =\ \frac{p_k (1 - (1\m t(f))^k)}{\sum_l p_l (1 - (1\m t(f))^l)},
\end{equation}
which is the expected observed degree distribution after covering fraction $f$ of nodes of graph $G$.

\smallskip
\subsubsection{Equivalence of traversal techniques under $RW(p_k)$}\label{sec:Equivalence_BFS_DFS}
An interesting observation is that, under the random graph model $RW(p_k)$, all common traversal techniques (BFS, DFS, FF, SBS, \ldots) are subject to exactly the same bias. This is because the sampled node sequence~$S$ is fully determined by the choice of stub indices on $[0,1]$, independently of the way we manage the elements in~$Q$.

This observation applies to the sequence~$S$ only - the subgraphs of $G$ that we actually sample by BFS and DFS, for example, might significantly differ.

\smallskip
\subsubsection{Equivalence to weighted sampling without replacement}\label{sec:Equivalence to weighted sampling}
Consider a node $v$ with a degree $k_v$. The probability that $v$ is discovered before time $t$, given that it has not been discovered before~$t_0\leq t$, is
\begin{equation}\label{eq:v_before_t}
    \Prob(\textrm{$v$ before time $t$ $|$ $v$ not before $t_0$}) = 1-\left(\frac{1\m t}{1\m t_0}\right)^{k_v}
\end{equation}
We now take a derivative $\frac{\textrm{d}}{\textrm{d} t}$ of the above equation, 
which results in the conditional probability density function $k_v(\frac{1\m t}{1\m t_0})^{k_v\m 1}$.
Setting $t\!\!\rightarrow\! t_0$ (but keeping $t\!>\!\!t_0$), reduces it to $k_v$, which is the density of probability that $v$ is sampled at $t_0$, given that it has not been sampled before.
This means that at every point in time, out of all nodes that have not yet been selected, the probability of selecting $v$ is proportional to its degree $k_v$. Therefore, this scheme is equivalent to  node sampling weighted by degree, without replacements. 

\smallskip
\subsubsection{Equivalence to RW for $f\!\!\rightarrow\! 0$}\label{sec:Equivalence to RW}
Finally, for $f\!\!\rightarrow\! 0$ (and thus $t\!\!\rightarrow\! 0$), we have $1\m (1\m t)^k \simeq k$, and Eq.~(\ref{eq:q_k_t}) simplifies to Eq.~(\ref{eq:q_k_RW}). This means that in the beginning of the sampling process, every traversal technique is equivalent to RW, as shown in Fig. 1 for $f\!\!\rightarrow\! 0$.

\smallskip
\subsubsection{$\langle k^*\rangle$ is decreasing in $f$} \label{subsec:k^* is decreasing in f}
Let us denote by $X_i\in V$ the $i$th selected node. As we have shown above that our procedure is equivalent to node degree weighted sampling without replacements, we can write:
\begin{eqnarray}
\nonumber  \Prob(X_1\eq u) &=& \frac{k_u}{z} \\
\nonumber  \Prob(X_2\eq w) &=& \sum_{u \neq w} \frac{k_{w}}{z-k_{u}}\cdot\frac{k_{u}}{z}\ = \frac{k_w}{z}\cdot \alpha_w,
\end{eqnarray}
where $z = 2|E|$ and $\alpha_{w} = \sum_{u \neq w} \frac{k_{u}}{z-k_{u}}$.
Because for any two nodes $a$ and  $b$, we have $\alpha_{b}\m \alpha_{a} = z(k_{a}\m k_{b}) / ((z\m k_{a})(z\m k_{b})),$
$\alpha_{w}$ strictly decreases with growing $k_{w}$. As a result, $\Prob(X_2)$ is more concentrated around nodes with smaller degrees than is $\Prob(X_1)$, implying that $\Mean[k_{X_2}] < \Mean[k_{X_1}]$. We can use an analogous argument at every iteration $i\leq |V|$, which allows us to say that $\Mean[k_{X_{i}}] < \Mean[k_{X_{i-1}}]$. In other words, $\langle k^*\rangle(f)$ is a decreasing function of $f$.

A practical consequence is that many short traversals (\emph{e.g.,} BFS-es) are more biased than a long one, with the same total number of samples.

\subsection{Comments on the starting node and graph connectivity}
In all exploration techniques, the choice of the starting node~$v_1$ can have a strong effect on the first iterations. For example, if $v_1$ is a low-degree node then the degree distribution $\est{q}_k$ sampled in the first iterations is naturally biased toward lower degrees. In fixed graphs, this problem is usually addressed by selecting $v_1$ as the last node of an appropriately long ``burn-in'' run of RW (or MHRW when this technique is used), started at an arbitrary node. In the case of a random graph $RG(p_k)$, the problem is even simpler, because already the second node of RW follows $\pi_v \eq\frac{k_v}{2|E|}$, which reduces the burn-in period to one hop only.


Another issue is that the configuration model $RG(p_k)$ might result in a graph $G$ that is not connected. In this case, every exploration technique covers only the component $C$ in which it was initiated; consequently, the process described in Section~\ref{subsec:Breaking the dependencies} stops once $C$ is covered.

\subsection{A convenient interpretation}
It might be sometimes convenient to split the exploration techniques, in $RG(p_k)$,  into three simple classes, with respect to the node degree bias they experience. These classes can be defined as ways to sample nodes from a pool of all nodes $V$, independently of the actual topology of $G$.
MHRW is equivalent to uniform node sampling with replacement.
RW is equivalent to degree-weighted node sampling with replacement.
Finally, all traversal techniques equivalent to degree-weighted node sampling without replacement.
The above holds strictly for $RG(p_k)$ only, but it can be an insightful interpretation, in general.

\section{Correcting for node degree bias} \label{sec:Correcting for node degree bias}
In the previous section we derived the expected observed degree distribution~$q_k$ as a function of the original degree distribution~$p_k$, for three general graph exploration techniques. The distribution~$q_k$ is usually biased towards high-degree nodes.
In this section, we derive unbiased estimators $\est{p}_k$ and $\langle\est{k}\rangle$ of the original degree distribution $p_k$ and its mean $\langle k\rangle$, respectively.

Let $S\subset V$ be a sequence of vertices that we sampled. Based on $S$, we can estimate $q_k$ as
\begin{eqnarray}
\label{eq:est{q}_k}   \est{q}_k &=&  \frac{\textrm{number of nodes in $S$ with degree $k$}}{|S|}  
\end{eqnarray}

\subsection{Random Walk (RW)}
In order to estimate $p_k$ based on $\est{q}_k$, consider again Eq.(\ref{eq:q_k_RW}), which says that $q_k$ is proportional to $k\,p_k$. Therefore,  $p_k$ is proportional to $q_k/k$, and $\est{p}_k$ is proportional to $\est{q}_k/k$ which allows us to write (similarly to \cite{Salganik04_RDS,Rasti09-RDS_Characterizing_Unstructured_Overlays}):
\begin{equation}\label{eq:est{p}_k_RW}
    \est{p}_k = \frac{\est{q}_k}{k}\ \cdot\ \left(\sum_l \frac{\est{q}_l}{l}\right)^{-1}
\end{equation}
where $\sum_l \frac{\est{q}_l}{l}$ is a normalizing constant.
From Eq.(\ref{eq:est{p}_k_RW}), we can estimate the average node degree as
\begin{equation}\label{eq:Ek_est_RW}
   \langle\est{k}\rangle\ =\ \sum_k k\,\est{p}_k \ =\  \left(\sum_l \frac{\est{q}_l}{l}\right)^{-1}=\frac{|S|}{\sum_{v\in S} \frac{1}{k_v}}
\end{equation}

\subsection{Metropolis Hastings Random Walk (MHRW)}
In this case, equations (\ref{eq:q_k_UNI}) and (\ref{eq:Ek_UNI}) trivially yield
\begin{eqnarray}
  \est{p}_k &=& \est{q}_k, \qquad \textrm{and} \\
\label{eq:est_k_UNI}    \langle\est{k}\rangle &=&  \sum_k k\,\est{p}_k  \ =\  \sum_k k\,\est{q}_k. 
\end{eqnarray}

\begin{figure*}
\psfrag{fraction f}[c][b][0.8]{$f$ - fraction of covered nodes}
\psfrag{degree}[c][t][0.8]{$\langle k^*\rangle$ - observed average node degree}
\psfrag{degree k}[c][b][0.9]{$k$ - node degree}
\psfrag{prob}[c][t][0.9]{$\Prob(k)$}
\psfrag{degreeDistr}[c][c][1]{Degree distribution}
\psfrag{zeroAss}[c][c][1]{Average node degree}
\psfrag{k1}[r][c][0.7]{$\langle k\rangle$}
\psfrag{k2}[r][c][0.7]{$\frac{\langle k^2\rangle}{\langle k\rangle}$}
\psfrag{L1}[l][c][0.55]{real, $p_k$}
\psfrag{L2}[l][c][0.55]{expected, $q_k$}
\psfrag{L3}[l][c][0.55]{RW, sampled, $\est{q}_k$}
\psfrag{L4}[l][c][0.55]{RW, estimate, $\est{p}_k$}
\psfrag{L5}[l][c][0.55]{BFS, $f\eq 0.1$, sampled, $\est{q}_k(f)$}
\psfrag{L6}[l][c][0.55]{BFS, $f\eq 0.1$, estimate, $\est{p}_k(f)$}
\psfrag{L7}[l][c][0.55]{BFS, $f\eq 0.3$, sampled, $\est{q}_k(f)$}
\psfrag{L8}[l][c][0.55]{BFS, $f\eq 0.3$, estimate, $\est{p}_k(f)$}
\includegraphics[width=1.\textwidth]{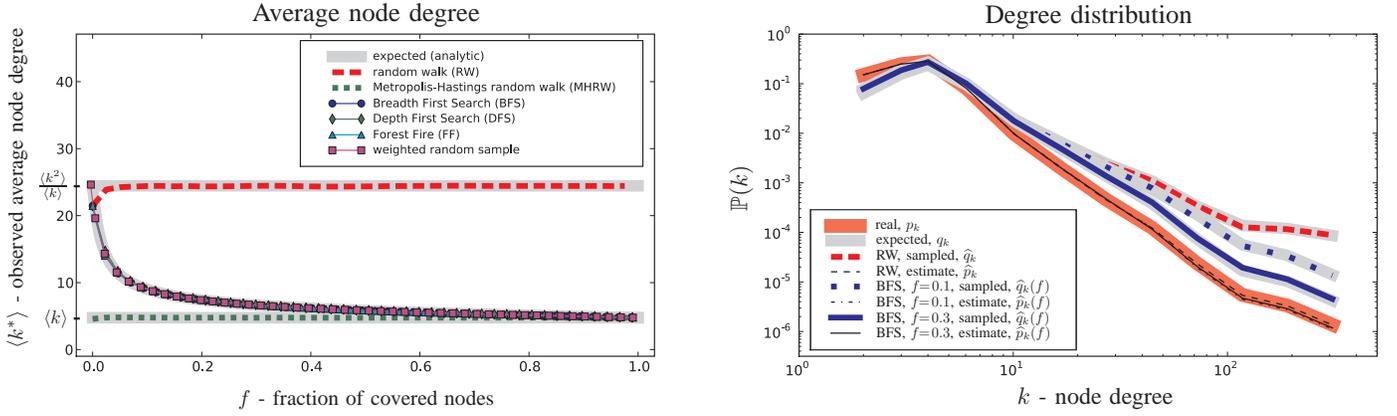}
\caption{{\bf Comparison of sampling techniques in theory and in simulation. }\textbf{Left:}  Observed (sampled) average node degree $\langle k^*\rangle$ as a function of the fraction $f$ of sampled nodes, for various sampling techniques.
The results are averaged over 1000 graphs with 10000 nodes each, generated by the configuration model with a fixed heavy-tailed  degree distribution~$p_k$ (shown on the right).
\quad \textbf{Right:} Real, expected, and estimated (corrected) degree distributions for selected techniques and values of $f$ (other techniques behave analogously).
\quad We obtained analogous results for other degree distributions and graph sizes $|V|$. The term
$\langle k\rangle$ is the real average node degree, and $\langle k^2\rangle$ is the real average squared node degree.
}
\label{fig:simulations.eps}
\end{figure*}

\begin{figure*}
\psfrag{fraction f}[c][b][0.8]{$f$ - fraction of covered nodes}
\psfrag{degree}[c][t][0.8]{$\langle k^*\rangle$ - average sampled node degree}
\psfrag{degree k}[c][b][0.9]{$k$ - node degree}
\psfrag{P(k)}[c][t][0.9]{$\Prob(k)$}
\psfrag{highAss}[c][c][1]{Average node degree, \ assortativity $r>0$}
\psfrag{lowAss}[c][c][1]{Average node degree, \ assortativity $r<0$}
\psfrag{k1}[r][c][0.7]{$\langle k\rangle$}
\psfrag{k2}[r][c][0.7]{$\frac{\langle k^2\rangle}{\langle k\rangle}$}
\includegraphics[width=1.\textwidth]{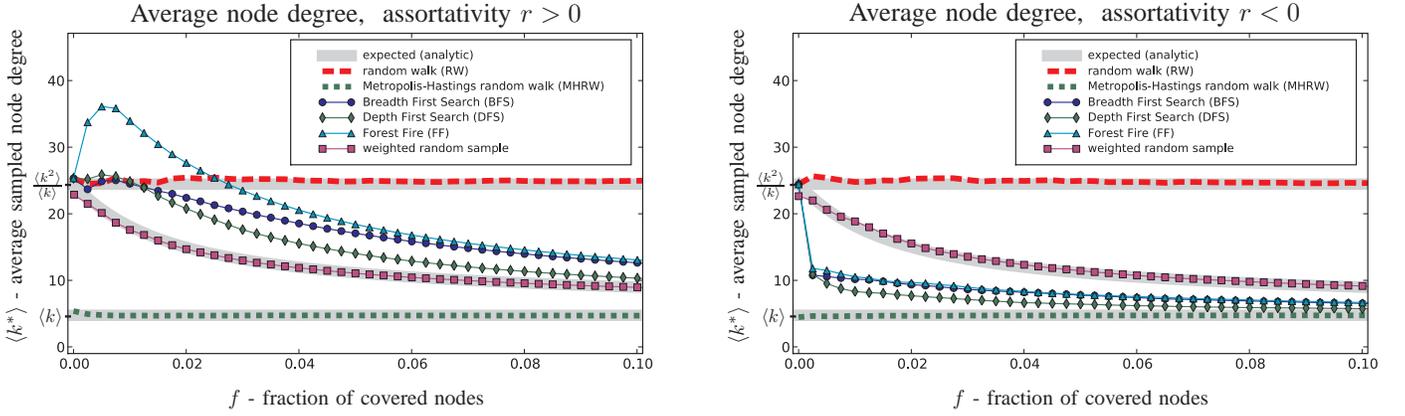}
\caption{{\bf The effect of assortativity $r$ on the results}. First, we use the configuration model with the same degree distribution~$p_k$ as in Fig.~\ref{fig:simulations.eps} (and the same number of nodes $|V|=10000$) to generate a graph $G$. Next, we apply the pairwise edge rewiring technique~\cite{Maslov02_science} to change the assortativity $r$ of $G$ without changing node degrees. This technique iteratively takes two random edges $\{v_1,w_1\}$ and $\{v_2,w_2\}$, and rewires them as $\{v_1,w_2\}$ and $\{v_2,w_1\}$ only if it brings us closer to the desired value of assortativity $r$. As a result, we obtain graphs with a positive (left) and negative (right) assortativity $r$. Note that for a better readability, we present only the values of $f\in[0,0.1]$, {\em i.e.,} ten times smaller than in Fig.~\ref{fig:simulations.eps}.
}
\label{fig:assortativity.eps}
\vspace{-0.5cm}
\end{figure*}

\subsection{Graph traversal}
From Eq. (\ref{eq:q_k_f}) we know that $p_k(f)$ is proportional to $q_k /(1 - (1\m t(f))^k)$. Consequently,
\begin{equation}\label{eq:est{p}_k_BFS}
    \est{p}_k\ =\ \frac{\est{q}_k}{1 - (1\m t(f))^k}\ \cdot\ \left(\sum_l \frac{\est{q}_l}{1 - (1\m t(f))^l}\right)^{-1}
\end{equation}
However, in order to evaluate this expression, we need to evaluate $t(f)$, that, in turn, requires $p_k$. We can solve this chicken-and-egg problem iteratively, if we know the real fraction $f^{real}$ of covered nodes, or equivalently the graph size $|V|$. First, we evaluate Eq.(\ref{eq:est{p}_k_BFS}) for some values of $t$ and feed the resulting $\est{p}_k$'s into Eq. (\ref{eq:f(t)}) to obtain the corresponding $f$'s. By repeating this process, we can drive the values of $f$ arbitrarily close to $f^{real}$, and thus find the desired $\est{p}_k$.

In summary, for graph traversal techniques, Eq.(\ref{eq:est{p}_k_BFS}) shows how to estimate the original degree distribution $p_k$ given that the real graph coverage $f^{real}$, which is often the case in practice. Of course, based on our estimator $\est{p}_k$, we can calculate the average node degree as $\langle\widehat{k}\rangle=\sum_k k\,\est{p}_k$.

\section{\label{sec:Simulation}Simulation results}

In this section, we implement and simulate the considered sampling techniques, namely BFS, DFS, FF (with $p=0.5$), RW and MHRW.
The simulations confirm our analytical results. More importantly, in simulations we can study the effect of topological properties, such as of assortativity, that are not directly captured by the random graph model~$RG(p_k)$.


\subsection{Estimating Degree Distributions and Average Degree}
Fig.~\ref{fig:simulations.eps} 
verifies all the formulae derived in this paper, for a random graph with a given powerlaw distribution.
 The analytical expectations are plotted in thick plain lines in the background and the averaged simulation results are plotted in thinner lines lying  on top of them. We observe almost a perfect match between  theory and simulation in estimating the sampled degree distribution $q_k$ (Fig.~\ref{fig:simulations.eps}, right) and its mean $\langle k^*\rangle$ (Fig.~\ref{fig:simulations.eps}, left). Indeed, all traversal techniques follow the same curve (as predicted in~\ref{sec:Equivalence_BFS_DFS}), that initially coincides with that of RW (see \ref{sec:Equivalence to RW}) and is monotonically decreasing in $f$
 (see \ref{subsec:k^* is decreasing in f}). We also show that degree weighted node sampling without replacements exhibits exactly the same bias (see \ref{sec:Equivalence to weighted sampling}). Finally, applying the estimators $\est{p}_k$ derived in Section~\ref{sec:Correcting for node degree bias} corrects for the bias of~$q_k$.


\subsection{The effect of degree-degree correlations (assortativity $r$)}
Depending on the type of network, nodes may tend to connect to similar or different nodes. For example, in most social networks high degree nodes tend to connect to other high degree nodes~\cite{Newman02}. Such networks are called \emph{assortative}. In contrast, biological and technological networks are typically \emph{disassortative}, {\em i.e.,} they exhibit significantly more high-degree-to-low-degree connections. This observation can be quantified by calculating the \emph{assortativity coefficient}~$r$~\cite{Newman02}, which is the correlation coefficient computed over all edges (\emph{i.e.,} degree-degree pairs) in the graph. Values $r\!<\!0$,  $r\!>\!0$ and $r\!=\!0$ indicate disassortative, assortative and purely random graphs, respectively.

For the same initial parameters as in Fig.~\ref{fig:simulations.eps} ($p_k$, $|V|$), we simulated different levels of assortativity.  Fig.~\ref{fig:assortativity.eps} shows the results.
Graph assortativity~$r$ strongly affects the first iterations of traversal techniques. Indeed, for assortativity $r>0$ (Fig.~\ref{fig:assortativity.eps}, left), the degree bias is even stronger than for $r=0$ (Fig.~\ref{fig:simulations.eps}, left). This is because the high-degree nodes are now interconnected more densely than in a purely random graph, and are thus easier to discover by sampling techniques that are inherently biased towards high degree nodes.
Interestingly, Forest Fire is by far the most affected. A possible explanation is that under Forest Fire, low-degree nodes are likely to be completely skipped by the first sampling wave.
Not surprisingly, a negative assortativity $r<0$ has the opposite effect: every high-degree node tends to connect to low-degree nodes, which significantly slows down the discovery of the former.

In contrast, random walks RW and MHRW are not affected by the changes in assortativity. This is expected, because their stationary distributions hold for \emph{any} fixed (connected and aperiodic) graph regardless of its topological properties.

\subsection{Other graph properties}
We also attempted to simulate the effect of other basic graph properties, such as clustering or modularity. However, all these properties are interdependent, which makes it difficult to interpret the results. For example, \cite{Newman09_RG_with_clustering}~described recently an extension of the configuration model to generate random graphs with a given level of clustering~$c$. However, the assortativity~$r$ turns out to strongly depend on~$c$. 
Rather than showing preliminary results, we decided to defer them to future work, where we are planning to incorporate some of these additional topological properties in our analytical model.


\begin{table}[t!]
  \centering
  {\footnotesize
\begin{tabular}{l|ccccc}
   & UNI              &   RW          &  BFS${}_{28}$   &  BFS${}_{1}$      & MHRW   \\
\hline
$|S|$              &  982K        & 2.26M & 28$\times$81K = 2.26M & 1.19M & 2.26M \\
$f$  &  0.44\%           & 1.03\%         &  28$\times$0.04\%     & 0.54\%  & 1.03\%
\end{tabular}}
  \caption{{\bf Facebook measurements - data set overview.} $|S|$ and $f$ are the absolute and relative lengths of the collected samples.
  For more details refer to~\cite{Gjoka09_Facebook_arxiv}.}
  \label{tab:Facebook_Datasets}\vspace{-0.8cm}
\end{table}

\section{\label{sec:Facebook}Real life example: Sampling of Facebook}
In this section we apply and test the previous ideas in a real-life large-scale system - the Facebook social graph.
With 250+ millions of active users, Facebook is currently the largest online social network. Crawling the entire topology of Facebook would require downloading about $50TB$ of HTML data~\cite{Gjoka09_Facebook_arxiv}, which makes sampling a very practical alternative.

\subsection{Data collection}
We have implemented a set of crawlers to collect the samples of Facebook (FB)
according to the UNI, BFS, RW, MHRW techniques.
The details of our implementation are described in~\cite{Gjoka09_Facebook_arxiv}.
The collected data sets are summarized in Table~\ref{tab:Facebook_Datasets}.

{\em UNI} refers to a uniform sample of FB users. It was obtained by uniformly sampling the entire FB userID space and discarding non allocated userIDs. This is a trivial version of rejection sampling and guarantees a uniform sampling of the existing users, regardless of their actual distribution in the userID space.
 UNI gives a high quality estimation of $p_k$ and $\langle k\rangle$, mainly thanks to a large number of samples~$|S|$. Therefore, we use UNI as ground truth for comparison of various techniques.

We ran two types of BFS crawling. BFS${}_{28}$ consists of 28 small BFS-es initiated at 28 randomly chosen nodes from UNI, which allowed us to easily parallelize the process. Moreover, at the time of data collection, we (naively) thought that this would reduce the BFS bias. After gaining more insight into the process (which, nota bene, motivated this paper), we collected a single large BFS${}_{1}$, initiated at a randomly chosen node from UNI.
The implementation of {\em RW} and {\em MHRW} is straightforward.

\begin{table}
  \centering
  {\footnotesize
\begin{tabular}{l|ccccc}
   & UNI & RW & BFS${}_{28}$ & BFS${}_{1}$ & MHRW \\
  \hline
  $\langle k^*\rangle$ sampled & 94.1 & 338.0 & 323.9 & 285.9 & 95.2 \\
  $\langle k^*\rangle$ expected & - & 329.8 (\ref{eq:Ek_RW}) & 329.1 (\ref{eq:q_k_f}) & 328.7 (\ref{eq:q_k_f}) & 94.1 (\ref{eq:Ek_UNI}) \\
  $\langle\est{k}\rangle$ estimated & - & 93.9 (\ref{eq:Ek_est_RW}) &  85.4 (\ref{eq:est{p}_k_BFS}) & 72.7 (\ref{eq:est{p}_k_BFS}) & 95.2 (\ref{eq:est_k_UNI})
\end{tabular}
}
\caption{{\bf Facebook measurements - average node degree.} Avg degree: sampled (row 1), expected (row 2) and corrected (row 3) for various techniques. For each expected and corrected value, we give in parenthesis the formula used to compute it.
}
\label{tab:Facebook_degrees}\vspace{-0.8cm}
\end{table}

\subsection{Results}
\begin{figure}[t]
\psfrag{degree k}[c][b][0.9]{$k$ - node degree}
\psfrag{prob}[c][t][0.9]{$\Prob(k)$}
\psfrag{degreeDistr}[c][t][1]{Degree distributions sampled in Facebook}
\center
\includegraphics[width=0.4\textwidth]{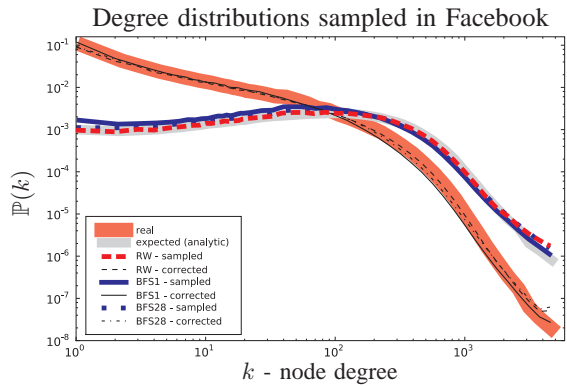}
\caption{{\bf Facebook measurements - degree distribution.} Crawlers used: UNI, RW and BFS. All plots are in log-log scale with logarithmic binning of data (we take the average of all points that fall in the same bin). We also correct these distributions, as described in Section~\ref{sec:Correcting for node degree bias}.}
\label{fig:Facebook_degree_disrtibutions}
\vspace{-0.6cm}
\end{figure}

We present the Facebook sampling results in Table~\ref{tab:Facebook_degrees} and in Fig.~\ref{fig:Facebook_degree_disrtibutions}. The first row of Table~\ref{tab:Facebook_degrees} shows the average node degree $\langle k^*\rangle$ observed (sampled) by several techniques. The value sampled by UNI is $\langle k^*\rangle\eq 94.1$, which we interpret as the real value~$\langle k\rangle$. MHRW, as expected, recovers a similar value. In contrast RW and BFS are both biased towards high degrees by a factor larger than three!
The degree bias of RW is the largest. It drops very slightly under the (relatively very short) BFS${}_{28}$ crawl, which confirms our findings from~\ref{sec:Equivalence to RW}.
BFS${}_{1}$, a sample 15 times longer than BFS${}_{28}$, is significantly less biased, which is in agreement with~\ref{subsec:k^* is decreasing in f}.

The second row shows the \emph{expected} sampled average node degrees ({\em i.e.,} our predictions of the values in the first row), assuming that the underlying Facebook topology is a random graph $RG(p_k)$ with degree distribution $p_k$ equal to that sampled by UNI. As expected, this works very well for RW. However, the values predicted for BFS significantly overshoot the reality. This is because Facebook \emph{is not} a random graph~$RG(p_k)$. For example, Facebook, as most social networks~\cite{Newman03_Review}, is characterized by a high clustering coefficient~$c$.
We believe that it is possible to incorporate this fact in our analytical model, \emph{e.g.,} by appropriately stretching the function $f(t)$ in Eq.~(\ref{eq:f(t)}). This is a main goal in our future work.

Finally, in the last row of Table~\ref{tab:Facebook_degrees} we apply the estimators developed in Section~\ref{sec:Correcting for node degree bias} to correct the degree biases of RW and BFS. In the case of RW, the correction works very well.
Unfortunately, for the BFS estimator the results are significantly worse, clearly for the reasons discussed in the previous paragraph.

All the above observations hold not only for the average node degree, but also for the entire degree distribution, which is shown in Fig.~\ref{fig:Facebook_degree_disrtibutions}.

\subsection{Practical recommendations}

BFS is strongly biased toward high degree nodes. It is possible to correct for this bias precisely when the underlying graph is a $RG(p_k)$ (which is not the case in practice). Also, in more realistic graphs, this bias can be corrected reasonably well for a very small sample size (as is the case for BFS${}_{28}$), where
BFS is similar to RW (see Fig.~\ref{fig:contributions}).
On the other extreme, for very large sampling coverage, the bias of BFS becomes relatively small and could be sometimes neglected (even without additional correction).
However, in all other cases, the results become difficult to interpret.
In contrast, both RW (equipped with a correction procedure) and MHRW are unbiased, regardless of the actual graph topology. Therefore, we recommend using RW and MHRW (with a slight advantage of RW~\cite{Rasti09-RDS_Characterizing_Unstructured_Overlays}) as general methods to sample the node properties.

In contrast, RW and MHRW are not really useful when sampling \emph{non-local graph properties}, such as the graph diameter or the average shortest path length. In this case, BFS seems very attractive, because it produces a full view of a particular region in the graph, which is usually a densely connected graph itself, and for which the non-local properties can be easily calculated. However, all such results should be interpreted very carefully, as they may be also strongly affected by the bias of BFS. For example, the graph diameter (usually) drops significantly with growing average node degree of a network.


%

\vspace{-0.1cm}
\section{\label{sec:Conclusion}Conclusion and Future Directions}
In this paper, we analyzed the bias in estimating node degree  when BFS (and other graph traversal techniques that sample nodes without replacement) are used to crawl a large, static, undirected network that is modeled by a random graph with a given, arbitrary degree distribution. We also compared BFS and graph traversal techniques to the well-studied random walks, and we were able to explain many of the similarities and differences that were only empirically observed so far. To the best of our knowledge, this is a first step towards analyzing the bias of BFS sampling, which is widely used in practice.
In future work, we plan to extend our theoretical framework and study the effect of topological properties other than the degree distribution (such as assortativity, clustering, or community structure)
on the bias of BFS and other techniques.

\bibliographystyle{IEEEtran}

\begin{thebibliography}{10}
\providecommand{\url}[1]{#1}
\csname url@samestyle\endcsname
\providecommand{\newblock}{\relax}
\providecommand{\bibinfo}[2]{#2}
\providecommand{\BIBentrySTDinterwordspacing}{\spaceskip=0pt\relax}
\providecommand{\BIBentryALTinterwordstretchfactor}{4}
\providecommand{\BIBentryALTinterwordspacing}{\spaceskip=\fontdimen2\font plus
\BIBentryALTinterwordstretchfactor\fontdimen3\font minus
  \fontdimen4\font\relax}
\providecommand{\BIBforeignlanguage}[2]{{%
\expandafter\ifx\csname l@#1\endcsname\relax
\typeout{** WARNING: IEEEtran.bst: No hyphenation pattern has been}%
\typeout{** loaded for the language `#1'. Using the pattern for}%
\typeout{** the default language instead.}%
\else
\language=\csname l@#1\endcsname
\fi
#2}}
\providecommand{\BIBdecl}{\relax}
\BIBdecl

\bibitem{monica}
M.~R. Henzinger, A.~Heydon, M.~Mitzenmacher, and M.~Najork, ``On near-uniform
  url sampling,'' in \emph{Proc. of WWW}, 2000.

\bibitem{willinger-unbiased-p2p}
D.~Stutzbach, R.~Rejaie, N.~Duffield, S.~Sen, and W.~Willinger, ``On unbiased
  sampling for unstructured peer-to-peer networks,'' in \emph{Proc. of IMC},
  2006.

\bibitem{Rasti09-RDS_Characterizing_Unstructured_Overlays}
A.~Rasti, M.~Torkjazi, R.~Rejaie, N.~Duffield, W.~Willinger, and D.~Stutzbach,
  ``Respondent-driven sampling for characterizing unstructured overlays,'' in
  \emph{INFOCOM Mini-Conference}, April 2009.

\bibitem{gkantsidis04randomwalksp2p}
C.~Gkantsidis, M.~Mihail, and A.~Saberi, ``Random walks in peer-to-peer
  networks,'' in \emph{Proc. of Infocom}, 2004.

\bibitem{Gjoka09_Facebook_arxiv}
M.~Gjoka, M.~Kurant, C.~T. Butts, and A.~Markopoulou, ``A walk in facebook:
  Uniform sampling of users in online social networks,''
  \emph{http://arxiv.org/abs/0906.0060}, 2009.

\bibitem{Twitter08}
B.~Krishnamurthy, P.~Gill, and M.~Arlitt, ``A few chirps about twitter,'' in
  \emph{Proc. of WOSN}, 2008.

\bibitem{Leskovec06_sampling-largegraphs}
J.~Leskovec and C.~Faloutsos, ``Sampling from large graphs,'' in \emph{Proc. of
  ACM SIGKDD}, 2006.

\bibitem{Lovasz93}
L.~Lovasz, ``Random walks on graphs. a survey,'' in \emph{Combinatorics}, 1993.

\bibitem{Najork01}
M.~Najork and J.~L. Wiener, ``Breadth-first search crawling yields high-quality
  pages,'' in \emph{Proc. of WWW}, 2001.

\bibitem{Ahn-WWW-07}
Y.~Ahn, S.~Han, H.~Kwak, S.~Moon, and H.~Jeong, ``{Analysis of Topological
  Characteristics of Huge Online Social Networking Services},'' in \emph{Proc.
  of WWW}, 2007.

\bibitem{Mislove-IMC-07}
A.~Mislove, M.~Marcon, K.~P. Gummadi, P.~Druschel, and S.~Bhattacharjee,
  ``{Measurement and Analysis of Online Social Networks},'' in \emph{Proc. of
  IMC}, 2007.

\bibitem{Wilson09}
C.~Wilson, B.~Boe, A.~Sala, K.~P. Puttaswamy, and B.~Y. Zhao, ``User
  interactions in social networks and their implications,'' in \emph{Proc. of
  EuroSys}, 2009.

\bibitem{Lee-Phys-Rev-06}
S.~H. Lee, P.-J. Kim, and H.~Jeong, ``Statistical properties of sampled
  networks,'' \emph{Phys. Rev. E}, vol.~73, p. 016102, 2006.

\bibitem{snowball-bias}
L.Becchetti, C.Castillo, D.Donato, and A.Fazzone, ``A comparison of sampling
  techniques for web graph characterization,'' in \emph{LinkKDD}, 2006.

\bibitem{MisloveWosn08}
A.~Mislove, H.~S. Koppula, K.~P. Gummadi, P.~Druschel, and B.~Bhattacharjee,
  ``Growth of the flickr social network,'' in \emph{Proc. of WOSN}, 2008.

\bibitem{Kim06_poisson_cloning}
J.~H. Kim, ``Poisson cloning model for random graphs,'' \emph{International
  Congress of Mathematicians (ICM)}, 2006 (preprint in 2004).

\bibitem{Achlioptas05_On_the_bias_of_traceroute_sampling}
D.~Achlioptas, A.~Clauset, D.~Kempe, and C.~Moore, ``On the bias of traceroute
  sampling: or, power-law degree distributions in regular graphs,'' in
  \emph{STOC}, 2005.

\bibitem{Shahbaz03_Sampling_with_Unequal_Probabilities}
M.~Q. Shahbaz, ``Sampling with unequal probabilities and without replacement,''
  Ph.D. dissertation.

\bibitem{Illenberger09_snowball_bias_correction}
J.~Illenberger, G.~Fl\"otter\"od, , and K.~Nage, ``An approach to correct bias
  induced by snowball sampling,'' \emph{Sunbelt Social Networks Conference},
  2009.

\bibitem{Goodman61_Snowball_sampling}
L.~Goodman, ``Snowball sampling,'' \emph{Annals of Mathematical Statistics},
  vol.~32, p. 148–170, 1961.

\bibitem{mcmc-book}
W.~Gilks, S.~Richardson, and D.~Spiegelhalter, \emph{Markov Chain Monte Carlo
  in Practice}.\hskip 1em plus 0.5em minus 0.4em\relax Chapman and Hall/CRC,
  1996.

\bibitem{Heckathorn97_RDS_introduction}
D.~Heckathorn, ``Respondent-driven sampling: A new approach to the study of
  hidden populations,'' \emph{Social Problems}, vol.~44, p. 174–199, 1997.

\bibitem{Salganik04_RDS}
M.~Salganik and D.~Heckathorn, ``Sampling and estimation in hidden populations
  using respondent-driven sampling,'' \emph{Sociological Methodology}, vol.~34,
  p. 193–239, 2004.

\bibitem{Leskovec05_Forest_Fire}
J.~Leskovec, J.~Kleinberg, and C.~Faloutsos, ``Graphs over time: densification
  laws, shrinking diameters and possible explanations,'' in \emph{KDD}, 2005.

\bibitem{Molloy95}
M.~Molloy and B.~Reed, ``A critical point for random graphs with a given degree
  sequence,'' pp. 161--179, 1995.

\bibitem{Newman03_Review}
M.~E.~J. Newman, ``The structure and function of complex networks,'' \emph{SIAM
  REVIEW}, vol.~45, pp. 167--256, 2003.

\bibitem{Newman01_EgoCentered_Networks}
------, ``Ego-centered networks and the ripple effect,'' \emph{Social
  Networks}, vol.~25, pp. 83--95, 2003.

\bibitem{Randomized_Algorithms_book}
R.~Motwani and P.~Raghavan, \emph{Randomized Algorithms}.\hskip 1em plus 0.5em
  minus 0.4em\relax Cambridge University Press, 1990.

\bibitem{Maslov02_science}
S.~Maslov and K.~Sneppen, ``Specificity and stability in topology of protein
  networks,'' \emph{Science}, vol. 296, no. 5569, pp. 910--913, May 2002.

\bibitem{Newman02}
M.~Newman, ``Assortative mixing in networks,'' in \emph{Phys. Rev. Lett. 89},
  2002.

\bibitem{Newman09_RG_with_clustering}
M.~E. J., ``Random graphs with clustering,'' \emph{Phys. Rev. Lett. (in
  press)}, 2009.

\end{thebibliography}


\end{document}